# The Eddington's Eclispe and a Possible Replica of the Experiment of Light Bending

C. Sigismondi

(1) ICRA/ICRANet Rome   sigismondi@icra.it

The success of the first measurement of the light bending by the solar gravitational field is due to the particular stellar field during the Eddington's 1919 total eclipse of the Sun, near the Hyades, giving the opportunity to measure the gravitational bending of the light to the astronomers in two expeditions in Brazil, Sobral, and on the Principe Island in the Atlantic Ocean. The geometrical properties of this field and another field in Leo are discussed in view of repeating this experiment of General Relativity with satellite data in the context of the International Year of Light 2015.

**The eclipse of Eddington: star field and instrumentation**

The occasion of verificating the predictions of the theory of General Relativity for the light bending arrived in 1919 with the total eclipse of May 29. The Sun was near the open cluster of Hyades rich of stars and suitable to check the deviation of the light around the mass of the Sun itself, the largest mass in the solar system. A maximum deviation of 1.75" is possible in case of stars grazing the solar limb. The deviation scales with the angular distance from the solar center as follows: $d\alpha = 1.75" * R_s/d"$ with $R_s = 959.63"$ at 1AU (Auwers, 1891).

During that eclipse the stars closer to the solar limb were at about 1 solar radius, with a theoretical deviation of $d\alpha < 0.87"$, that with the telescope of 4 m of focal length (the 4" astrograph[1]) used in the observations gave a difference of $5800*\tan(0.87")$ mm~25 µm in the same star field observed during the night 6 months after the eclipse. A long debate has been raised among the historians on the validity of the measurements (*there are cases in astronomy when the correct result is known the observations are beyond the possibility of the instruments*[2]) when the stellar Airy disks enlarged by the atmospheric turbulence are larger than the displacements to be found. Re-analyses of the data showed agreement with the results published in 1920 by Dyson, Eddington and Davidson.[3]

**SOHO Coronographs**

The Satellite SOHO Solar Heliospheric Observatory is in orbit around the Lagrangian point L1 since February 1996 and it gives daily images of the Sun since then in different wavelengths and with two coronographs LASCO (Large Aperture Spectrometric Coronograph) C2 and LASCO C3 for studying the space climate, through the solar corona and the coronal mass ejections observed in white light. The two coronographs have different field of view, C2 narrower than C3, respectively about 3° and 17°. The optical scheme of C2 is in the figure after.[4]

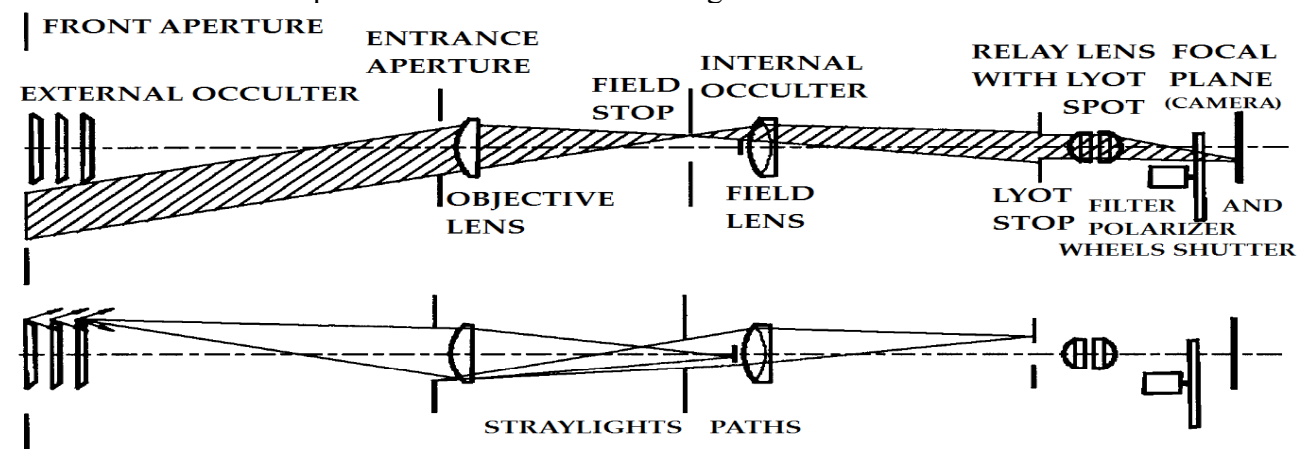

---

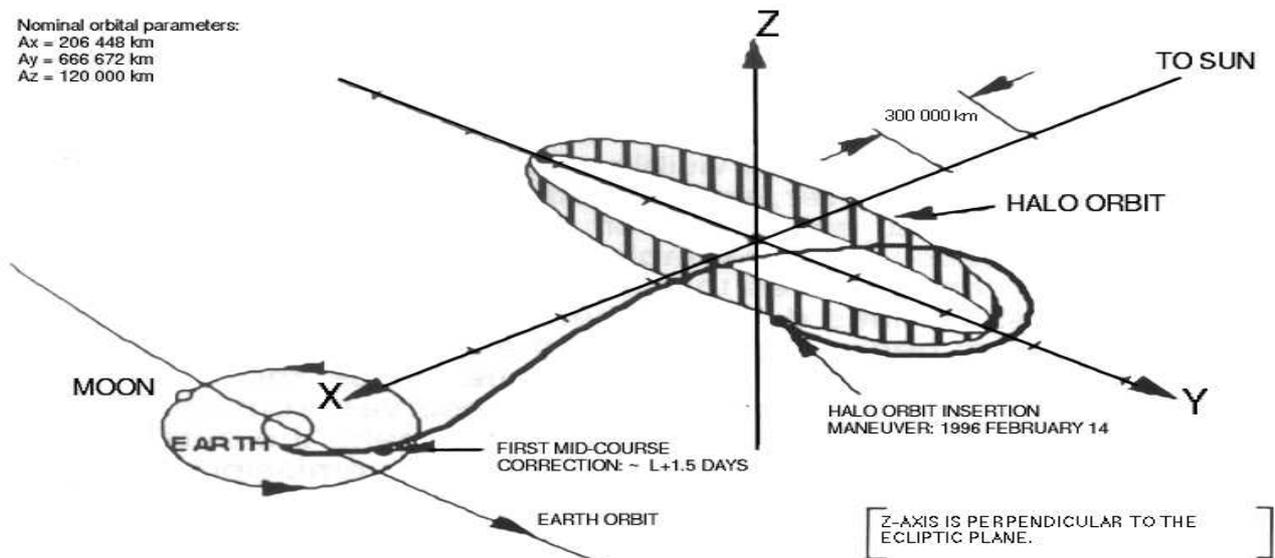

SOHO orbit schematic

The orbit of SOHO is an halo orbit of 179 days of period,[5] and 666672 Km x 202448 Km in the plane X,Y of the Ecliptic and 120000 Km in the plane perpendicular to the Ecliptic. The Lagrangian point L1 is at 1500000 Km from the Earth (0.01 AU) and this geometry avoided in 2004 June 8 and 2012 June 6 to observe the transits of Venus from SOHO, because with Venus at 0.2887 AU from the Earth and the spacecraft 120000 off the plane of the ecliptic (near the line Sun-Earth where the orbit reaches the largest distance from the ecliptic) the angle of displacement of Venus on the solar surface was 573" with respect to the 945"-544"=401" at maximum transit in 2012 and 945"-635"=310" in 2004 (945" was the solar diameter as seen from Earth and 544"or 635" the distances of Venus from the solar center during the transits).

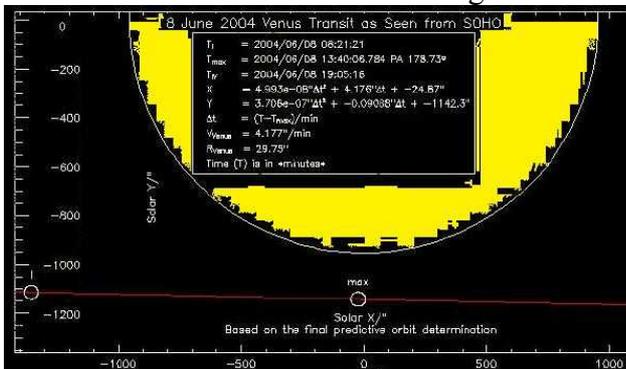
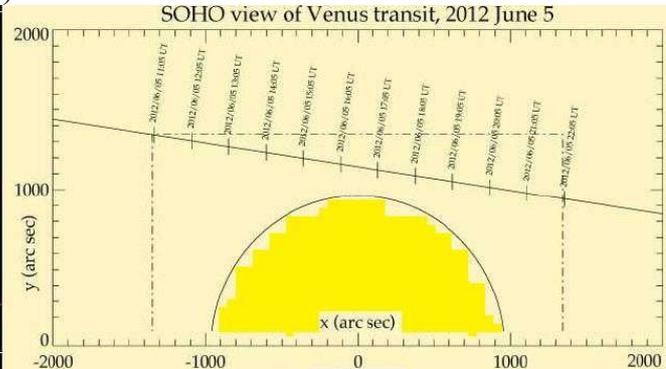

SOHO missed transits of Venus: in 2004 SOHO was higher and in 2012 lower than the ecliptic plane, 16.31 orbits after: the fraction of orbit made after the 16th put SOHO on the opposite side of the plane, loosing again the alignment to see the transit.

**The field of view of SOHO LASCO C2 coronograph**

The C2 coronograph is the more suitable for detecting the effect of the deviation due to the solar gravitational field. The occulter excludes the view of the stars below 1 solar radius from the solar limb. I have chosen the days from 21 to 24 august 2014, during wich Regulus, α Leo, and 31 Leo with ν Leo appear in the field of view of the instrument. Their V magnitudes are 1.4, 4.35 and 5,25. They form a triangle which is deformed by the optical distortions of the instrument as they progress in the field of view. α and ν are aligned almost horizontally and α and 31 are rather vertical in the HIRES 1024x1024 pixels square images, taken each 12 minutes. The first measurement of the field of view has been made with the drift-scan of Regulus from left to right: 1024 px per 7 pixel in vertical from 08/21 at 5.4 h UT to 08/24 at 19 UT; from ephemerides (Stellarium 12.4) the

---

5   A. Graps, http://solar-center.stanford.edu/FAQ/QL1.html  also  http://soho.nascom.nasa.gov/explore/faq.html
C. S. Roseberry, http://ccar.colorado.edu/asen5050/projects/projects_2003/roseberry/  and finally C. S. Roberts, http://www.ieec.cat/hosted/web-libpoint/papers/roberts.pdf Figure: http://sohowww.nascom.nasa.gov/gif/halo_orbit.gif

Sun covered 206.11 arcmin along the ecliptic so the scale length ratio is 12.076"/px.

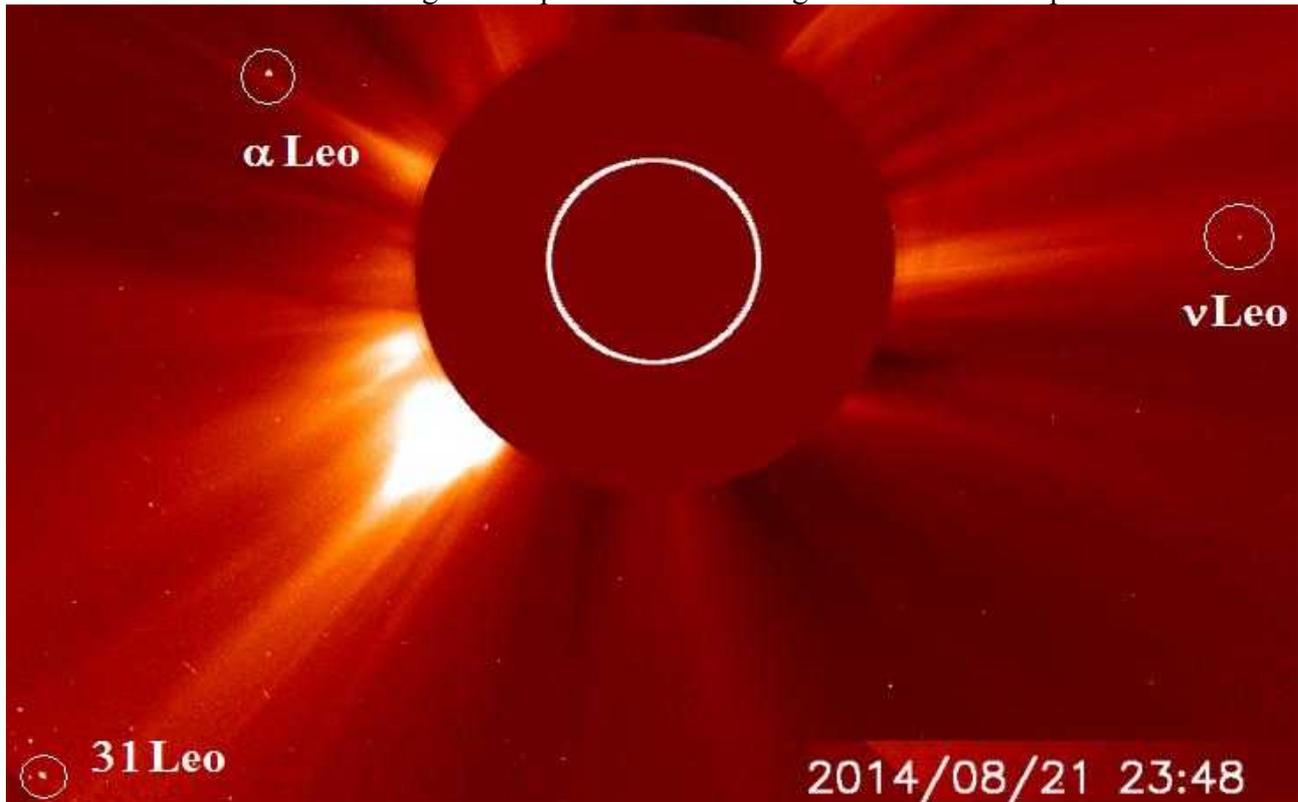

Recovering the distances α-ν 119'.57±0'.60 and α-31 153'.41±0'.32, both are larger than the distances calculated from stellar coordinates of 118'.3 and 151'.55, a systematic difference of 1.6'±0'.3 corresponding to 8±1 pixels, so well measurable, it is due to the spacecraft's motion along its orbit, going in the direction opposite to the Earth's orbit during these 85.6 hours. The sides of the triangle α-ν-31, measured in pixels, change slightly across the field of view. This is due to the optical deformation of the instrument.

**Discussion of the data, conclusions and perspectives**

The selection of Regulus has been made because it is on the ecliptic plane,[6] and the star almost grazes the Sun.

The variations of the sides of the triangle α-ν-31 Leonis as measured with SOHO C2 images give the optical distortions's field: in the vertical direction the distortion is larger than in the horizontal direction. Another source of distortion is the external occulter with the diffraction around it: the point-spread function of the star elongates itself toward the radial direction as it approaches the occulter's rim.[7] The amplitude of the effect at 1 solar radius from the solar limb, available to C2 coronograph, is less than 0.9", say 7/100 of a pixel. Hundred images with Regulus enough close to the occulter's rim are required to reduce statistically the error of a factor of 10. But in 1200 min, almost one day the star moves of 1°, two solar radii, so the study has to be made over all 19 transits of Regulus during the 19 years of operations. The long focal length lenses used in the 1919 eclipse and the continuity of the photographic emulsion granted a space resolution better than 11.5"/pixel[8] and a final errorbar of 0.11" hundred times smaller than the one obtained from a single SOHO image. The optical distortion has been evalued as radial distortion $\delta\rho(mm)=\rho m-\rho=a\rho+b\rho^3$: $\rho$ is the real distance from the optical center and $\rho m$ is the measured distance, $a=5.4\cdot10^{-3}$ and $b=-1.3\cdot10^{-5}$ it is a classical barrel distortion up to 5 pixels at the borders of the image.[8]

---

6  As ρ Leo chosen to calibrate SOHO magnitudes along the field of view (from pixels 0 to 320 see fig. http://lasco-www.nrl.navy.mil/content/level_1/c2/vig/rholeo_c2.east.gif while from pixels 710 to 1024 see http://lasco-www.nrl.navy.mil/content/level_1/c2/vig/rholeo_c2.west.gif . In the intermediate pixels the Sun and the external occulter hide the star. Regulus is not occulted by the Sun itself but only by the rim of the occulter, ν Leo from both.

7  All the instrumental calibrations before and after the launch are published in the websites of SOHO, but here we replicated all measurements using the bright stars approaching the solar limb and visible from C2 coronograph.

8  Llebaria, A., Ph. Lamy and S. Plunkett, Astronomical Data Analysis Software and Systems VI, 125, 435 (1997)

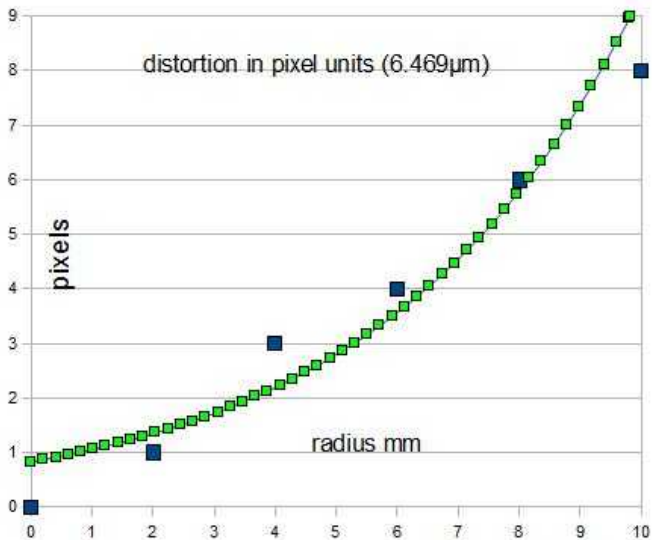
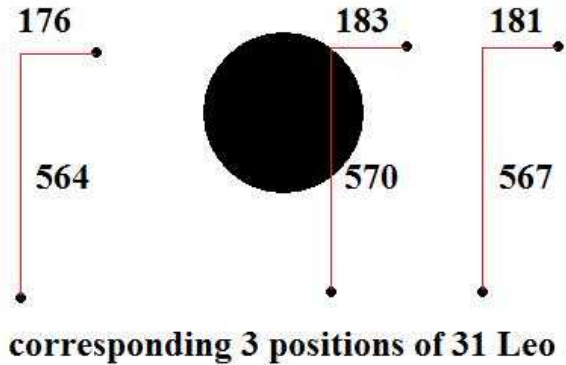

The three positions of Regulus are respectively at pixel (226,368) (715,370) and (995,374), pixel (0,0) is upper left. The vertical distance reaches a maximum near the center of the image, stretched by 3 to 6 pixels/570 as well as the horizontal dimension, stretched by 2 to 7 pixels/180. The left part of the figure is the radial distortion according to Llebaria et al., 1997, the right part shows our measurement from 22 to 24 Aug 2014. The black circle on the right is the 2.4 mm disk at the center of the field of view. The radial distortions in pixels are computed from the table in Llebaria et al. 97. An evenly spaced reticulum, according to these measurements, is transformed into a barrel-like pattern. There is a difference between the left function and our measurements: for the central position of the two stars the stretching either in vertical and horizontal is the maximul, instead of being the minimum (since the distance from the optical center is minimal). This is explained by the diffraction generated by the occulting disk. Four diffraction lines are very well visible and spaced of $\Delta x=8$ pixels each, 92" or 516μm: with nominal focal lenght F=364 mm $\Delta x = F \cdot \theta = F \cdot \lambda / D$, $\lambda$=550 nm. After the radial correction the accuracy on the position become a pixel, but we are still far from the required 1/20 pixel necessary to detect the light bending. Moreover the PSF diameter of the stars is about 9 pixels for Regulus and 5 pixels for 31 Leo, when they are far from the central disk, where for Regulus it becomes an 10x2 pixels ellipse, due to the occulter's diffraction. These distortions prevent to reduce to 1/20 pixel the errorbar, unless using more than 400 stacked images along 20 yr.